\documentclass[%
 reprint,
 amsmath,amssymb,
 aps, nofootinbib
]{revtex4-1}

\usepackage{graphicx}
\usepackage{dcolumn}
\usepackage{bm}
\usepackage{lipsum}
\usepackage[utf8]{inputenc}
\usepackage{verbatim}

\begin{document}

\title{Properties of the Innermost Stable Circular Orbit of a spinning particle moving in a rotating Maxwell-dilaton black hole background}

\author{Carlos Conde}
 \email{cacondeo@unal.edu.co}
\affiliation{Universidad Nacional de Colombia. Sede Bogot\'a. Facultad de Ciencias. Observatorio Astronómico Nacional. Ciudad Universitaria. Bogota, Colombia.
}

\author{Cristian Galvis}
 \homepage{cagalvisf@unal.edu.co}
\affiliation{
Universidad Nacional de Colombia. Sede Bogot\'a. Facultad de Ciencias. Departamento de F\'isica. Ciudad Universitaria. Bogota, Colombia.
}%
\author{Eduard Larra\~{n}aga}
 \homepage{ealarranaga@unal.edu.co}
\affiliation{%
 Universidad Nacional de Colombia. Sede Bogot\'a. Facultad de Ciencias. Observatorio Astronómico Nacional. Ciudad Universitaria. Bogota, Colombia.}%

\date{\today}

\begin{abstract}
 In this paper we investigate the innermost stable circular orbit (ISCO) of a spinning test particle moving in the rotating Maxwell-dilaton black hole spacetime. By using the Mathisson-Papapetrou-Dixon equations along with the Tulczyjew spin-supplementary condition, we find the equations of motion in the equatorial plane and, from the radial equation, it is obtained the effective potential for the description of the particle's motion. The obtained trajectories show that the ISCO radii for spinning particles moving in rotating charged backgrounds are always smaller than those obtained in the corresponding Kerr-Newman spacetimes. The increasing in the particle's spin produces a decrease in the ISCO radius in all the studied cases, with a maximum value for the spin and a corresponding minimum ISCO radius, obtained by imposing a condition that guarantees the timelike nature of the particle's worldline.

\end{abstract}


\maketitle

\section{Introduction}

One of the most interesting properties of the orbits of massive particles in general relativity is that there is a minimal radius at which they may have stable circular motion around a compact object such as black holes. This innermost stable circular orbit (ISCO) is defined by the geometrical properties of spacetime, i.e. the background metric, but it may be also corrected by the spinning properties of the moving particle. One of the first studies in this area was made by Suzuki and Maeda \cite{Suzuki1998} in which they considered the equatorial motion of a classical spinning particles in both Schwarzschild and Kerr metrics in the "pole-dipole" approximation, obtaining the properties of the ISCO numerically. Later, Jefremov et. al. \cite{Jefremov2015} studied the same motion to find the corrections to the ISCO radius using the approximation of small spin and recently, Zhang et. al. \cite{Zhang2018} studied the same configuration, imposing the additional condition that the velocity of the spinning particle must remain timelike.

The apparition of the ISCO is not exclusive of general relativity. In fact many geometric theories of gravity with solutions representing compact objects such as black holes have orbits of particles with similar properties. A special case corresponds to low-energy effective field theory describing string theory, which have been shown to contain black hole solutions with properties that are qualitatively different from those that appear in general relativity, but share some of the orbit motion properties. In this work we will analyze the circular motion of a spinning particle moving around the Sen black hole \cite{Sen1992} which is an exact solution in the low-energy effective heterotic string theory representing a compact object with a finite amount of electric charge and angular momentum. This solution generalizes the charged static metric known as the GMGHS black hole \cite{Gibbons1982, Gibbons1988, Garfinkle1991, Garfinkle1992}. 

The work in this paper is organized as follows. In Section 2 we give a brief review of the rotating Maxwell-dilaton charged black hole. In Section 3 we obtain the equations of motion for a spinning test particle in this black hole background. For this purpose, we adopt the tetrad formalism to find the four-momentum components by using the conserved quantities equations associated with the corresponding Killing vectors in the tetrad frame. Afterward, we find the four-momentum components in the coordinate frame and based on both Riemann and spin tensors calculation, we get  the equations of motion in the equatorial plane. In Section 4 we define the effective potential of the spinning test particle and compute numerically the Innermost Stable Circular Orbit (ISCO) and impose an additional constraint to ensure that the velocity of the particle is always timelike. The solution of these equations let us obtain the ISCO properties in terms of the spin of the test particle. Finally, some conclusions are given in Section 5.

\vspace{-0.5cm}

\section{The Kerr-Sen black hole}

The Einstein-Maxwell-Dilaton rotating black hole solution that will be considered in this work arises from the action of low energy heterotic string theory in four dimensions, which can be written in the string frame as \cite{Sen1992, Yazadjiev2000} 
\begin{align}
\nonumber
 \mathcal{S} &= \int d^4 x \sqrt{- g^{(s)}} e^{-\Phi} \\[1ex]
  & \ \ \times  \left[ R + 4 \partial^\mu \varphi \partial_\mu \varphi - \frac{1}{8} F^2 - \frac{1}{12} H^2 + \Lambda  \right]
\end{align}
where $R$ is the scalar curvature, $\Phi$ is the dilaton field, the Maxwell field $ A^\mu$ enters in the term  $F^2 = F^{\mu \nu} F_{\mu \nu}$ with $F_{\mu \nu} = \partial_\mu A_\nu - \partial_\nu A_\mu$, $\Lambda$ is the cosmological constant and the factor $H^2 = H_{\mu \nu \sigma}  H^{\mu \nu \sigma}$ represents a tensor field related with a 2-form potential $B_{\alpha \beta}$ and the Maxwell field by $H_{\mu \nu \sigma}  = \partial_\mu  B_{\nu \sigma} + \partial_\nu  B_{\sigma \mu} + \partial_\sigma  B_{\mu \nu } - \frac{1}{4} \left( A_\mu  F_{\nu \sigma} + A_\nu  F_{\sigma \mu} + A_\sigma  F_{\mu \nu } \right)$. The stringy effects are mediated by the factor $e^{-\Phi}$ and therefore we obtain the Einstein metric, through the conformal transformation $g_{\mu \nu} =e^{-\Phi} g^{(s)}_{\mu \nu}$.\\  
Sen \cite{Sen1992, Yazadjiev2000} found a charged, stationary, axially symmetric solution of the field equations resulting from the above action in the Einstein frame by using target space duality applied to the Kerr solution. This solution has a vanishing field $H$ and no cosmological constant. Its line element, in generalized Boyer-Linquist coordinates, is 

\begin{eqnarray}
ds^{2} & = & -\left(\frac{\Delta - a^2 \sin^2 \theta}{\Sigma}\right)dt^{2} +  \frac{\Sigma}{\Delta} dr^{2} +\Sigma d\theta^{2} \nonumber \\
 &  & -\frac{4\mu ar \cosh^2 \alpha \sin^{2}\theta}{\Sigma}dtd \phi +\frac{\Xi}{\Sigma}\sin^{2}\theta d\phi^{2},\label{eq:kerrsen2}
\end{eqnarray}

where 

\begin{eqnarray}
\Delta & = & r^2 - 2 \mu r +a^{2}\\[1ex]
\Sigma & = & r^2 +a^{2}\cos^{2}\theta +2 \mu r \sinh ^2 \alpha \\[1ex]
\Xi & = & ( r^2 + 2\mu r \sinh ^2 \alpha + a^2 )^2 - a^2 \Delta \sin ^2 \theta .
\end{eqnarray}

The constant $a=\frac{J_{BH}}{M}$ is identified with the specific angular momentum of the black hole while the constants $\alpha$ and $\mu$ in this expression are related with the black hole's mass, $M$, and  electric charge, $Q$, through the relations
\begin{align}
\mu= M - \frac{Q^2}{2M} \\
\sinh ^2 \alpha = \frac{Q^2}{2M^2 - Q^2}
\end{align}
or 
\begin{align}
M = \mu \cosh ^2 \alpha \ \ \ \ \ Q = \frac{\mu}{\sqrt{2}} \sinh 2\alpha \ .
\end{align}

The Kerr-Sen's solution reduces to the static charged GMGHS black hole \cite{Gibbons1982,Gibbons1988,Garfinkle1991,Garfinkle1992} when $a=0$ and becomes Kerr's solution when $\alpha=0$ (or equivalently, $Q=0$).
This metric has an event horizon given by the biggest root of the equation $\Delta=0$. Its radius is written in terms of the black hole parameters as

\begin{equation}
r_{H}=M-\frac{Q^{2}}{2M}+\sqrt{\left(M-\frac{Q^{2}}{2M}\right)^{2}-a^{2}} \ .\label{eq:horizon}
\end{equation}

Note that, from equation (\ref{eq:horizon}), the horizon disappears unless
\begin{equation}
\left|a\right|\leq M -\frac{Q^{2}}{2M},
\end{equation}
and therefore, there is the possibility of an extremal black hole when $Q^2 = 2M(M-a)$ or equivalently $\mu = a$. 
For further calculations, we mention here the non-vanishing components of the inverse metric for the Kerr-Sen solution,

\begin{equation}
\begin{aligned}
g^{tt}&=- \dfrac{ \Xi \Sigma}{\Delta\Xi + a^{2}\sin^{2}\theta \left( 4r^{2}\mu^{2}\cosh^{4}\alpha-\Xi \right)}\, , \\[1ex]
g^{t\phi} &= - \dfrac{ 2a\mu r \cosh^{2}\alpha \Sigma}{\Delta\Xi + a^{2}\sin^{2}\theta \left( 4r^{2}\mu^{2}\cosh^{4}\alpha-\Xi \right)}\, , \\[1ex]
g^{rr} &= \dfrac{\Delta}{\Sigma} \, , \ \ \ \ \ \  \ g^{\theta \theta} = \dfrac{1}{\Sigma}\, , \\[1ex]
g^{\phi \phi} &= \dfrac{(\Delta\csc^{2}\theta-a^{2})\Sigma}{\Delta\Xi + a^{2}\sin^{2}\theta \left( 4r^{2}\mu^{2}\cosh^{4}\alpha-\Xi \right)} \ \ . 
\end{aligned}
\label{eq:inverse}
\end{equation}

\section{Equations of Motion for a Spinning Particle}

The motion of a spinning test particle in the background of the Kerr-Sen black hole are given by the Mathisson-Papapetrou-Dixon equations \cite{Mathison1937, Papapetrou1951, Dixon1970},
\begin{align}
\label{eq:eom1}
\frac{DP^\mu }{D\tau} &= -\frac{1}{2} R^\mu _{ \ \nu \rho \sigma} u^\nu S^{\rho \sigma} \\[2ex]
\label{eq:eom2}
\frac{DS^{\mu \nu} }{D\tau} &= P^\mu u^\nu- P^\nu u^\mu,
\end{align}
where 
\begin{equation}
u^\mu = \frac{dx^\mu}{d\tau}
\end{equation}
is the 4-velocity of the test particle while $P^\mu$ is its 4-momentum and $ R^\mu _{ \ \nu \rho \sigma}$ is the Riemann tensor of the Kerr-Sen metric. The antisymmetric tensor
$S^{\mu \nu} = - S^{\nu \mu}$ 
is the spin tensor of the test particle, which gives its spinning angular momentum, $s$, through the relation
\begin{equation}
s^2 = \frac{1}{2} S^{\mu \nu} S_{\mu \nu}.
\label{eq:s}
\end{equation}

There is also a freedom in the choice of a spin-supplementary condition which may produce different trajectories \cite{Zhang2018}. In this work, we will choose the Tulczyjew spin-supplementary condition,
\begin{equation}
P_\mu S^{\mu \nu} = 0,
\label{eq:Tulczyjew}
\end{equation}
which restricts the spin tensor to generate only rotations \cite{Hojman2013}. It is also important to note that, although the momentum vector of the orbiting particle satisfies the normalization condition $P_\mu P^\mu = - m^2$,
the particle's velocity may not. In fact, when the orbiting particle is spinning, its momentum vector is not always parallel to its velocity vector and therefore the later may be timelike or spacelike along the trajectory.

In order to obtain the equations of motion for a spinning test particle in the Kerr-Sen spacetime, we proceed by defining a local orthonormal tetrad such that the metric in Eq. \eqref{eq:kerrsen2} writes
\begin{equation}
g_{\mu \nu} = \eta_{(a)(b)}e^{(a)}_{\mu}e^{(b)}_{\nu} \ \ ,
\label{eq:Kerr-Sen}
\end{equation}
where the latin indices label the tetrad. Hence, the tetrad components are
\begin{align}
\nonumber
e^{(0)}_{\mu}dx^{\mu} &= \sqrt{\dfrac{\Delta}{\Sigma}} \left(  dt - a\sin^{2}\theta d\phi  \right)\\[1ex] \nonumber
e^{(1)}_{\mu}dx^{\mu}&= \sqrt{\dfrac{\Sigma}{\Delta}} \, dr  \\[1ex]
e^{(2)}_{\mu}dx^{\mu} &= \sqrt{\Sigma} \, d\theta \\[1ex] \nonumber
e^{(3)}_{\mu}dx^{\mu} &= \dfrac{a\sin\theta}{\sqrt{\Sigma}} \Big[ -a dt + \left( r^{2} + 2\mu r \sinh^{2}\alpha + a^{2} \right) d\phi  \, \Big] \ ,
\label{corrected1}
\end{align}
while the inverse elements are obtained using the relation
\begin{equation}
e^{\mu}_{(a)} = \eta_{(a)(b)}e^{(b)}_{\ \nu}g^{\mu \nu} \ \ .
\label{eq:dual}
\end{equation} 
giving
\begin{widetext}
\begin{align}
\nonumber
e^{\mu}_{(0)}\partial_{\mu} &= \dfrac{\sqrt{\Delta\Sigma} \left( \Xi - 2a^{2}r\mu\cosh^{2}\alpha \sin^{2}\theta \right)}{\Delta\Xi + a^{2}\sin^{2}\theta \left( 4r^{2}\mu^{2}\cosh^{4}\alpha - \Xi \right)} \partial_{t}  - \ \dfrac{\sqrt{\Delta\Sigma}a \left( a^{2}\sin^{2}\theta - 2r\mu \cosh^{2}\alpha - \Delta  \right)  }{\Delta\Xi + a^{2}\sin^{2}\theta \left( 4r^{2}\mu^{2}\cosh^{4}\alpha - \Xi \right)} \partial_{\phi} \\[2ex] 
e^{\mu}_{(1)} \partial_{\mu} &= \sqrt{\dfrac{\Delta}{\Sigma}} \partial_{r} \\[2ex] \nonumber
e^{\mu}_{(2)}\partial_{\mu} &= \dfrac{1}{\sqrt{\Sigma}} \partial_{\theta} \\[2ex] \nonumber
e^{\mu}_{(3)}\partial_{\mu} &= - \dfrac{a\left( 2a^{2}\mu r \cosh^{2}\alpha + 2\mu r^{3}\cosh^{2}\alpha + 4\mu^{2}r^{2}\sinh^{2}\alpha \cosh^{2}\alpha - \Xi \right)\sqrt{\Sigma}\sin{\theta}}{\Delta\Xi + a^{2}\sin^{2}\theta \left( 4r^{2}\mu^{2}\cosh^{4}\alpha - \Xi \right)} \partial_{t} \\[1ex] \nonumber
& \ \ \ \ + \dfrac{\sqrt{\Sigma}\csc\theta \big\{ \Delta \left[ a^{2} + r\left( r + 2\mu \sinh^{2}\alpha \right) \right] \csc^{2}\theta - a^{2}\left[ a^{2} + r(r-2\mu) \right] \big\}}{\Delta \Xi \csc^{2}\theta + a^{2}\left( 4r^{2}\mu^{2}\cosh^{2}\alpha  - \Xi\right)}\partial_{\phi}
\label{eq:corrected2}
\end{align}
\end{widetext}

If the test particle moves on the equatorial plane, equations \eqref{eq:eom1} and \eqref{eq:eom2} imply that the only nonvanishing component of the spin vector is $s^{(2)} = -s$ \cite{Saijo1998}. Therefore, the nonvanishing components of the spin tensor in the tetrad frame are,
\begin{equation}
\begin{aligned}
\begin{cases}
S^{(0)(1)} &= -sP^{(3)} =- S^{(1)(0)}  \\[0.5ex]
S^{(0)(3)} &= s P^{(1)} = -S^{(3)(0)} \\[0.5ex]
S^{(1)(3)} &= sP^{(0)} = - S^{(3)(1)}
\end{cases}
\end{aligned}
\end{equation}

Note that in the Kerr-Sen black hole there are two Killing vectors, $\xi^{\mu}  \equiv ( 1 ,0,0, 0)$ and $ \phi^{\mu}   \equiv ( 0,0,0,1)$. In the tetrad frame, the components of these vectors are 
\begin{align}
 \xi^{(a)} & \equiv \left( \, \sqrt{\dfrac{\Delta}{\Sigma}} \, , \ 0 \, , \, 0 \, , \, -\dfrac{a\sin \theta}{\sqrt{\Sigma}} \, \right)\, ,  \\[1ex]  \nonumber
 \phi^{(a)} &\equiv \left( -a \sqrt{\dfrac{\Delta}{\Sigma}}\sin^{2} \theta \, , \, 0 \, , \, 0 \, , \, \dfrac{\sin \theta}{\sqrt{\Sigma}} \sqrt{ \Xi + a^{2}\Delta \sin^{2}\theta }   \right).
\end{align}

For a spinning test particle, the conserved quantity corresponding to a given Killing vector $\zeta$ is
\begin{equation}
C_{\zeta} = P^{\mu}\zeta_{\mu} - \dfrac{1}{2}S^{\mu \nu} \nabla_{\mu}\zeta_{\nu} .
\label{eq:conserved}
\end{equation}

Using \eqref{eq:conserved} in the tetrad frame and using the antisymmetry relation of the spin tensor together with the Killing equation $\nabla_{(\mu)}\zeta_{(\nu)}=-\nabla_{(\nu)}\zeta_{(\mu)}$, the conserved quantities associated with the Killing vectors $\xi$ and $\phi$ are the energy $C_{\xi}=E$ and the angular momentum $C_{\phi}=J$. The equations for these constants of motion are,
\begin{align}
\label{eq:Econserved}
- E&=  -P^{(0)}\xi^{(0)} + P^{(3)}\xi^{(3)} +  sP^{(3)} \nabla_{(0)} \xi_{(1)}\, , \\[2ex] \nonumber
J &= - \phi^{(0)}P^{(0)} + \phi^{(3)}P^{(3)} + sP^{(3)} \nabla_{(0)} \phi_{(1)} \\[1ex] & \hspace{0.4cm}- s P^{(0)} \nabla_{(1)}\phi_{(3)} \ \ ,
\label{eq:Jconserved}
\end{align}

where the nonvanishing components of the covariant derivatives of the Killing vectors in the tetrad frame are given by\footnote{Equations for the tetrad components and the covariant derivatives of the Killing vectors are given here to correct some typos appearing in the work of An et. al. \cite{An2018}.}
\begin{align}
\nonumber
 \nabla_{(0)}\xi_{(1)} &= - \dfrac{\mu \cosh^{2}\alpha\left( a^{2}\cos^{2}\theta - r^{2} \right)}{\Sigma^{2}}\, , \\[2ex] \nonumber
\nabla_{(0)}\phi_{(1)} &= -a\sin\theta \Big[ r\left( 2 \left(r+\mu \sinh^{2}\alpha \right)^{2}- r^{2} + \mu r\right) \\ & \ \ \ \,  + a^{2}\cos^{2}\theta(r-\mu) \Big] \Big/ \Sigma^{2} \, , \\[2ex] \nonumber
\nabla_{(1)}\phi_{(3)} &=  \dfrac{\sqrt{\Delta} \sin \theta \left(  r + \mu\sinh{2}\alpha \right)}{\Sigma^{2}} \ . 
\label{corrected3}
\end{align}

In order to study the circular motion of the spinning particle, we will consider equatorial motion, i.e. $\theta = \frac{\pi}{2}$, and therefore $P^{(2)}=P^{(\theta)}=0$. Using the conservation equations \eqref{eq:Econserved}-\eqref{eq:Jconserved} and the relation $P^{(a)}P_{(a)}=-m^{2}$, the components of the 4-momentum of the spinning particle in the tetrad frame are
\begin{equation}
\begin{aligned}
P^{(0)} &= \dfrac{\Sigma_{0}}{\sqrt{\Delta}} \dfrac{X}{Z}\, , \hspace{1cm} P^{(1)} = \pm \dfrac{\sqrt{\mathcal{R}}}{\sqrt{\Delta}Z}\, , \\[2ex]
P^{(2)} &= 0\, , \hspace{1.85cm} P^{(3)} = \Sigma_{0}^{2}\dfrac{Y}{Z} \, , \\[2ex]
\end{aligned}
\end{equation}
where
\begin{equation}
\begin{aligned}
X &= \left[ \sqrt{\Sigma_0^5} + \sqrt{\Sigma_0^3} a^2 + s a r F \right] E \\[0.5ex] & \ \ - \left[ s \mu r^2 \cosh ^2 \alpha + \sqrt{\Sigma_0^3} a \right] J \\[2ex]
Y &= \sqrt{\Sigma_0} J - \left[ s(r+\mu r \sinh ^2 \alpha) + \sqrt{\Sigma_0} a \right] E  \\[2ex]
Z &= \Sigma_0^3 - s^2 \mu r^2 \cosh ^2 \alpha (r + \mu \sinh ^2 \alpha) \\ 
\end{aligned}
\label{eq:X-Y-Z_functions}
\end{equation}
with 
\begin{equation}
\begin{aligned}
\Sigma_0 &=  r^2 +2 \mu r \sinh ^2 \alpha \\[2ex]
F &= r^2 +4 \mu r \sinh^2 \alpha + 2\mu^2 \sinh^4 \alpha + \mu r
\end{aligned}
\end{equation}
and 
\begin{equation}
\mathcal{R} = \Sigma_0^2 X^2 - \Sigma_0^4 Y^2 \Delta - m^2 Z^2 \Delta \ \ .
\end{equation}

The 4-momentum components in the coordinate frame are obtained through the projection $P^{\mu} = e^{\mu}_{(a)} P^{(a)}$, giving \begin{align}
\nonumber
P^0 &=  \frac{\sqrt{\Sigma_0^3}}{Z \left[ \Delta \Xi_0 +a^2 (4\mu^2 r^2 \cosh ^4 \alpha - \Xi_0) \right] } \nonumber  \\
& \times \left\{ \left( \Xi_0 -2 \mu a^2 r \cosh ^2 \alpha \right) X \right. \nonumber \\ \nonumber
 &  \ \ \ +\left[ \Xi_0 -2 \mu r \cosh ^2 \alpha (a^2+r^2 + 2\mu r \sinh^{2}\alpha) \right]  a \Sigma_0 Y \big\}\, , \\[2ex] 
P^1 &=  \pm \frac{\sqrt{\mathcal{R}}}{\sqrt{\Sigma_{0}} Z}\, , \hspace{1cm} P^2   =    0 \, , \\[2ex] \nonumber
P^3 &=  \frac{\sqrt{\Sigma_0^3}}{Z \left[ \Delta \Xi_0 +a^2 (4\mu^2 r^2 \cosh ^4 \alpha - \Xi_0) \right] } \nonumber  \\
 & \times \left\{ \left( \Delta +2 \mu r \cosh ^2 \alpha +a^2 \right) a X \right. \nonumber \\ \nonumber
 & \left. \ \ +\left( \Delta [a^2 +r ( r+ 2 \mu \sinh ^2 \alpha)] - a^2 [a^2+ r(r-2\mu )] \right)  \Sigma_0 Y \right\} \ .
\end{align}

Using the Tulczyjew spin-supplementary condition \eqref{eq:Tulczyjew}, the components of the spin tensor are
\begin{align}
S^{01} = -\dfrac{S^{31}P_{3}}{P_{0}} \ \ , \hspace{1.5cm} S^{03} = \dfrac{S^{31}P_{1}}{P_{0}} \ ,
\label{eq:Spin4-Momentum}
\end{align}
and after expanding equation \eqref{eq:s} and replacing \eqref{eq:Spin4-Momentum} we obtain,
\begin{equation}
S^{31} = \dfrac{sP_{0}}{m} \sqrt{\dfrac{g_{22}}{-g}}
\end{equation}
where $g = \textrm{det} ( g_{\mu \nu} )$. Therefore, the nonvanishing spin tensor components in the Kerr-Sen background can be written as 
\begin{equation}
\begin{aligned}
\begin{cases}
S^{01} &= -S^{01} = \dfrac{sP_{3}}{m \sqrt{\Sigma_{0}}} \\[2ex]
S^{03} &= -S^{30} = - \dfrac{sP_{1}}{m\sqrt{\Sigma_{0}}} \\[2ex]
S^{31} &= -S^{13} = \dfrac{sP_{0}}{m \sqrt{\Sigma_{0}}} \ .
\end{cases}
\end{aligned}
\end{equation}

The equations of motion in \eqref{eq:eom1}-\eqref{eq:eom2} can be expressed in terms of the spin tensor components as
\begin{align}
\nonumber
P^{0}\dot{r}-P^{1} &= \dfrac{1}{2}\dfrac{s}{m\sqrt{\Sigma_{0}}} g_{3\mu}R^{\mu}_{\ \nu \alpha \beta}u^{\nu}S^{\alpha \beta} \\[1ex]
& \ \ \ \, + \dfrac{sP_{3}(r + \mu \sinh^{2}\alpha)}{m\Sigma_{0}^{3/2}} \dot{r}  \\[2ex] \nonumber
P^{0}\dot{\phi} - P^{3} &= -\dfrac{sP_{1}}{m}\dfrac{(r+\mu\sinh^{2}\alpha)}{\Sigma_{0}^{3/2}}\dot{r} \\[1ex]
& \ \ \ \, - \dfrac{1}{2} \dfrac{s}{m \sqrt{\Sigma_{0}}}g_{1\mu}R^{\mu}_{\ \nu \alpha \beta}u^{\nu}S^{\alpha \beta} 
\end{align}

The nonvanishing components of the Riemann tensor relevant in these equations are presented in the appendix. Substituting these values, the nonzero components of the 4-velocity are
\begin{align}
\nonumber
\dot{r} &= P^{1}\left[ 1 + \dfrac{s^{2}g_{11}}{m^{2}\Sigma_{0}} R_{3003} \right] \bigg[ P^{0}  + \dfrac{s^{2}}{m^{2}\Sigma_{0}}R_{3113}P_{0} \\[2ex] \label{eq:EOM1}
& \hspace{0.8cm}  + \dfrac{s^{2}}{m^{2}\Sigma_{0}}R_{3101}P_{3} - \dfrac{s}{m} \dfrac{r+\mu \sinh^{2}\alpha}{\Sigma^{3/2}}P_{3}\bigg]^{-1} \\[4ex] \nonumber
\end{align}

\begin{align}
\nonumber
\dot{\phi} &= \bigg[P^{3} + \dfrac{s^{2}}{m^{2}\Sigma_{0}} \left( R_{1001}P_{3} + R_{1013}P_{0} \right) \\[2ex]  \nonumber
& \ \ \ \ \ \ \  - \dfrac{sP_{1}}{m \sqrt{\Sigma_{0}}} \dfrac{r + \mu \sinh^{2}\alpha}{\Sigma_{0}} \dot{r} \bigg] \\[2ex]
& \ \ \ \, \times \left[ P^{0} - \dfrac{s^{2}}{m^{2} \Sigma_{0}} \left(R_{1301}P_{3} + R_{1313}P_{0} \right) \right]^{-1} \ .
\label{eq:EOM2}
\end{align}

\section{ISCO of a Spinning Particle moving around the Kerr-Sen Black Hole}

In order to determine the ISCO for a spinning test particle in the Kerr-Sen geometry, we consider the square of $P^{1}$, which can be factorized, introducing the specific energy and angular momentum of the particle, $e=\frac{E}{m}$ and $j=\frac{J}{m}$, as
\begin{align}
\nonumber
(P^{1})^{2} &=\frac{m^{2}}{\Sigma_{0}Z^{2}}\left(A e^{2} + B e + C\right)\\ \nonumber
	&= \frac{m^{2}A}{\Sigma_{0} Z^{2}} \left(e-\frac{-B+\sqrt{B^{2}-4AC}}{2A}\right)\\
	& \hspace{0.5cm}\times\left(e+\frac{B+\sqrt{B^{2}-4AC}}{2A}\right)
	\label{eq:P1_momentum}	
\end{align}
where A, B and C are:
\begin{eqnarray}
A &=& \Sigma_0^2 \left[ K_1^2 - \Delta \Sigma_0^2 K_3^2 \right] \\
B &=& 2 \Sigma_0^2  j \left[ K_1 K_2 - \Delta \Sigma_0^2 K_3 K_4 \right] \\
C &=& \Sigma_0^2 j^2 \left[ K_2^2 - \Delta \Sigma_0^2 K_4^2 \right] - Z^2 \Delta
\end{eqnarray}
and $K_{1}$, $K_{2}$, $K_{3}$, $K_{4}$ are defined as
\begin{eqnarray}
K_1 &=& \Sigma_0^{5/2} + \Sigma_0^{3/2} a^2 -sarF\\
K_2 &=& s\mu r^2 \cosh^2 \alpha - \Sigma_0^{3/2} a\\
K_3 &=& s (r + \mu \sinh^2 \alpha ) - \Sigma_0^{1/2} a\\
K_4 &=& \Sigma_0^{1/2}. 
\end{eqnarray}

Since $\dot{r}$ is parallel to $P^{1}$, the condition for circular orbits, $\dot{r} = 0$, implies that $P^{1} = 0$ and therefore it is possible to define an effective potential for the spinning test particle in the Kerr-Sen spacetime as  
\begin{equation}
V_{\textrm{eff}} =  \frac{-B + \sqrt{B^2 - 4 A C}}{2 A}.
\label{eq:V-effective}
\end{equation}

\begin{figure}[!htb]
    \centering
    \includegraphics[width=1\linewidth]{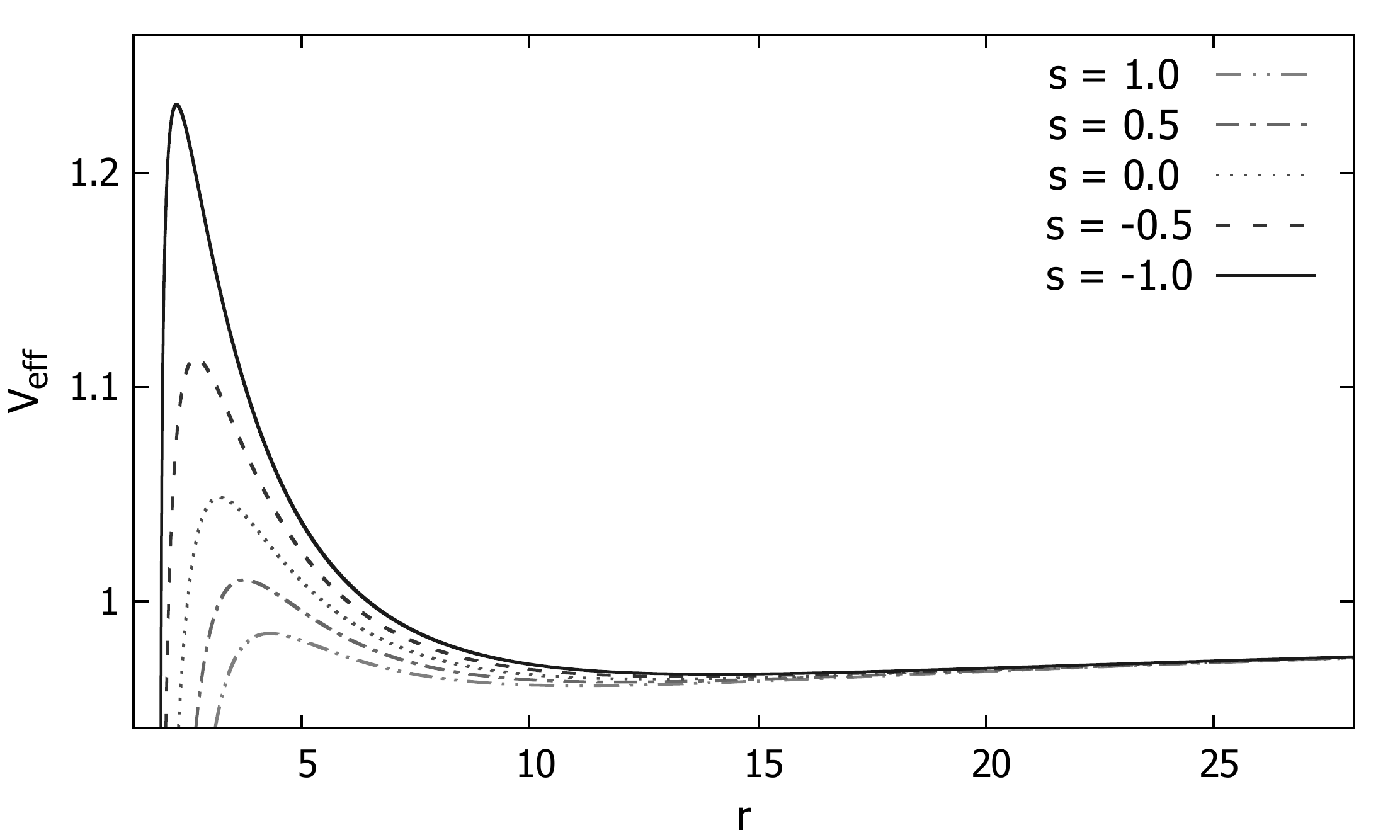}
    \caption{Effective potential for a spinning test particle with different spin in Kerr-Sen geometry with parameters set as $l = 4$, $a=0.25$ and $Q=0.25$ }
    \label{fig:V-effective}
\end{figure}

The typical behaviour of this effective potential is shown in Figure \ref{fig:V-effective}. The chosen parameters are such that the spinning test particle has angular momentum $l=4$ and the black hole has  $a=0.25$ and $Q=0.25$ and the curves correspond to different values of the particle’s spin, $s$.

Using equation \eqref{eq:V-effective} together with the conditions
\begin{align}
\frac{dV_{\textrm{eff}}}{dr} = 0 \ \  \hspace{1.5cm} \frac{d^2 V_{\textrm{eff}}}{dr^2} = 0 \ ,
\end{align}
it is possible to obtain the parameters $s$, $e$ and $j$ defining the ISCO for the spinning test particle. However, there is one more thing to take into account. As we mention above, since the orbiting particle is spinning, its momentum vector $P^\mu$ is not always parallel to its velocity $u^\mu$. Therefore, in order to ensure the motion of the test particle to be physical, we will also impose a superluminal constraint,
\begin{equation}
    \frac{u^2}{(u^0)^2} = \frac{u^\mu u_\mu}{(u^0)^2} = g_{00} + g_{11}\dot{r}^2+ 2 g_{03} \dot{\phi}  + g_{33} \dot{\phi}^2  < 0.
    \label{eq:superluminal}
\end{equation}

\subsection{Spinning Particle in Schwarzschild's Background}
\begin{figure}[htb!]
	\centering
	\includegraphics[width=1 \linewidth]{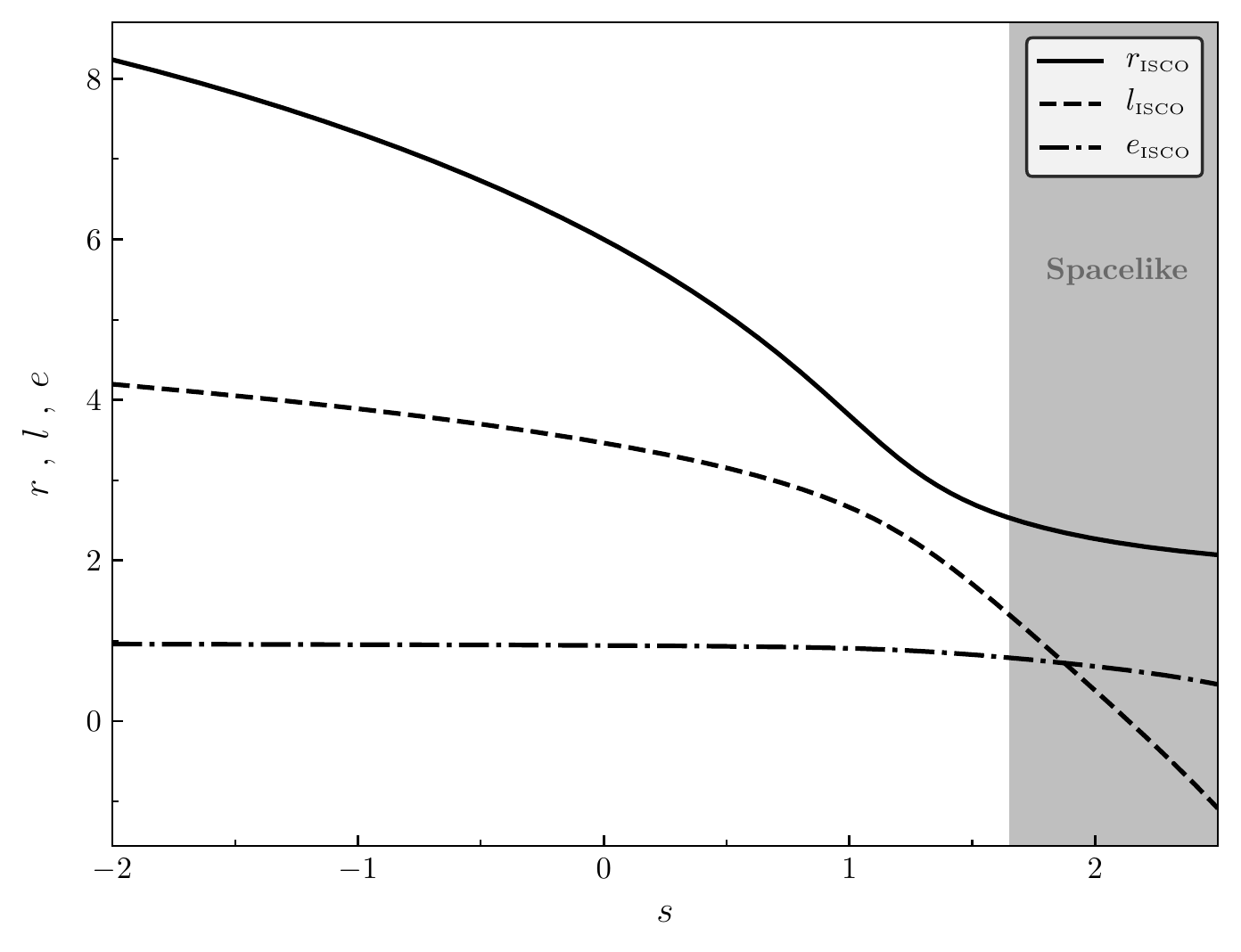}
	\caption{ISCO parameters in terms of the spin $s$ for a spinning test particle moving in Schwarzschild’s geometry.}
	\label{fig:ISCO-Schwarzschild}
\end{figure}

Setting $a=0$ and $Q=0$ in equation (\ref{eq:kerrsen2}) it is recovered the Schwarzschild black hole. By solving numerically the ISCO equations for different values of $s$, it is obtained the orbit radius $r_{\textrm{ISCO}}$, the specific angular momentum $l_{\textrm{ISCO}}$, and the specific energy $e_{\textrm{ISCO}}$, as shown in Figure \ref{fig:ISCO-Schwarzschild}. The vertical line at $s \approx 1,6518$ divide the plot in two parts according to the norm of the velocity vector; for values in the unshaded side, the velocity is timelike and the motion is physically possible, while values of the parameters in the shaded side of the plot give a spacelike velocity vector and hence, a non-physical motion. 

Note that non-spinning particles, $s=0$, move at an ISCO  with the well known radius $r_{\textrm{ISCO}} = 6M$. As the particle's spin increases, the ISCO radius decreases, reproducing the results in \cite{Jefremov2015} and \cite{Zhang2018}. This result can be understood noting that, when the particle has some positive value of spin, it adds an intrinsic contribution to the total angular momentum,  allowing it to get closer to the compact object. This fact is stressed in the curve of the specific angular momentum, which also decreases as the spin increases. Similarly, the specific energy $e$ also decreases as the spin gets larger.\\

In Table \ref{tab:ISCO-Schwarzschild} the values of the ISCO parameters, together with the values of the angular velocity, $\Omega^{\frac{2}{3}}$, and the norm of the velocity vector, $u^{2}$, are presented for different values of spin. From these, it is clear that the angular speed increases with the spin because the radius of the ISCO decreases and that the value of $u^{2}$ decreases, getting closer to zero (i.e. the particle increases its speed, approaching the speed of light), becoming first a light-like vector and after a spacelike vector in the gray region. These results reproduce those obtained by Zhang et. al. \cite{Zhang2018}. \\
The values of the ISCO parameters at the point at which $u^2 = 0$ are


\begin{eqnarray*}
    r_{\textrm{ISCO}} \approx 2.5299 \hspace{2.8em} l_{\textrm{ISCO}} \approx 1.3227\\
    e_{\textrm{ISCO}} \approx 0.7894 \hspace{2em} \ s_{\textrm{max}} \approx 1.6518 \, .
\end{eqnarray*}

\begin{table}[htb!]
	\centering
	\begin{tabular}{cccccc}
		\hline
		$s$    & $r_{\textrm{ISCO}}$    & $l_{\textrm{ISCO}}$    & $e_{\textrm{ISCO}}$    & $\Omega^{\frac{2}{3}}$ & $u^2 $ \\ \hline \hline
		-0.9 	& 	7.2135	    & 3.8542	& 	0.9528 	& 	0.1455 	& -0.5624  \\
		-0.7 	&	6.9807 	& 3.7792 	& 	0.9512 	& 	0.149 		& -0.5523 \\
		-0.5 	&	6.7294	    & 3.6985	&	0.9492	& 	0.1530 	& -0.5405 \\
		-0.3 	&	6.4568	    & 3.6111	&	0.9470	& 	0.1578 	& -0.5265\\
		-0.1 	&	6.1594		& 3.5156	& 	0.9443	& 	0.1634 	& -0.5097\\
		0    	&	6.0			& 3.4641	&	0.9428	&	0.1667 	& -0.5 \\
		0.1  	&	5.8325	    & 3.4097	&	0.9411	&	0.1703 	& -0.4892\\
		0.3  	& 	5.4699	    & 3.2906	&	0.9371	&	0.1787 	& -0.4637\\
		0.5  	&	5.0633	    & 3.1533	&	0.9319	&	0.1894 	& -0.4308\\
		0.7  	& 	4.6028		& 2.9901	&	0.9248	&	0.2035 	& -0.3870 \\
		0.9  	&	4.0834	    & 2.7871	&	0.9143	&	0.2224 	& -0.3267 \\
		\hline 
	\end{tabular}
	\caption{Parameters of the ISCO for a spinning particle moving in the Schwarzschild geometry ($a=0$ and $Q=0$).}
	\label{tab:ISCO-Schwarzschild}
\end{table}

\vspace{-0.5cm}

\subsection{Spinning Particle moving in a GMGHS Background}

\begin{figure}[htb!]
	\centering
	\includegraphics[width=1 \linewidth]{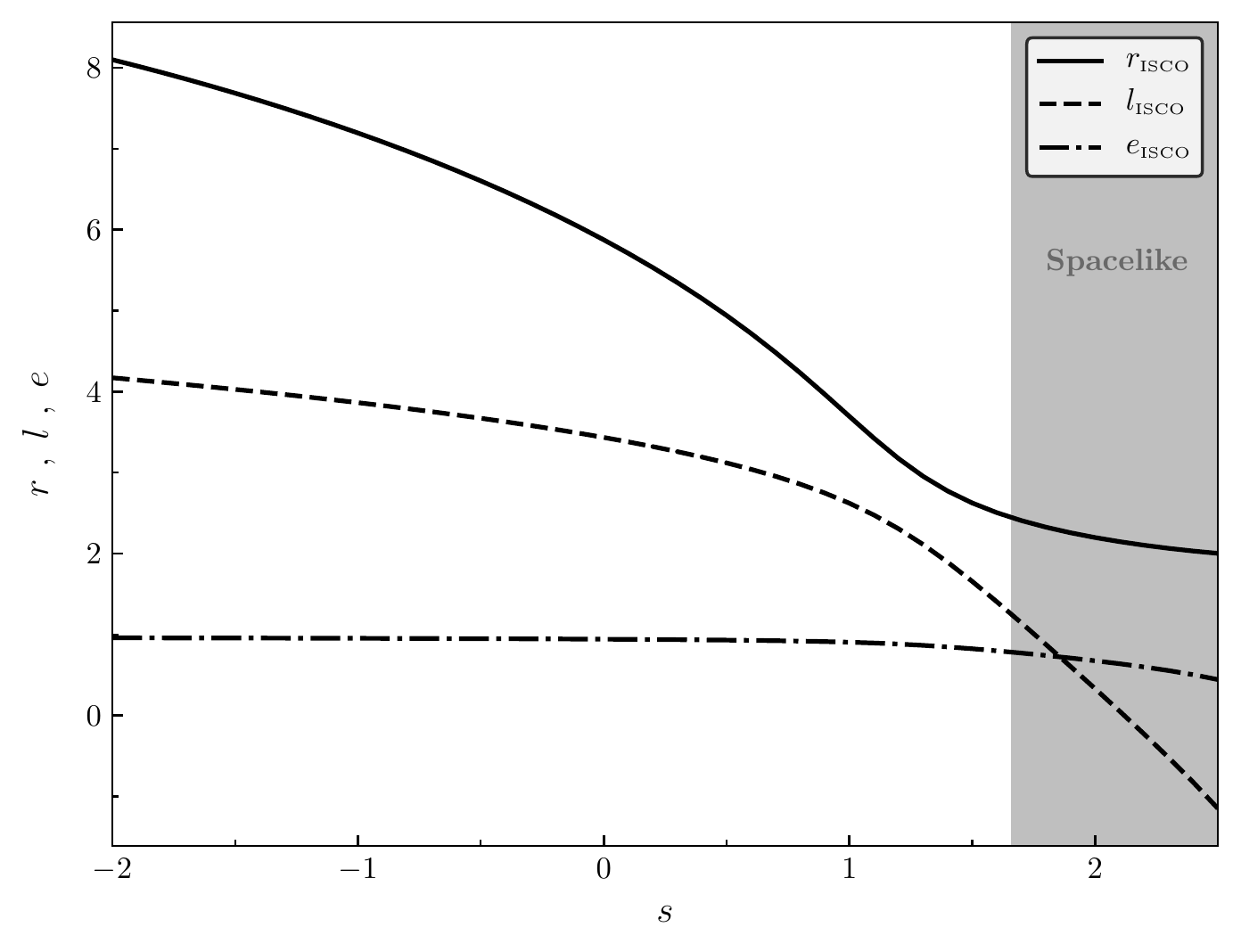}
	\caption{ISCO parameters in terms of the spin $s$ for a spinning test particle moving in the GMGHS background with $Q=0.25$.} 
	\label{fig:ISCOparametersQ}
\end{figure}

The GMGHS black hole solution \cite{Gibbons1982, Gibbons1988, Garfinkle1991, Garfinkle1992} is obtained from the metric (\ref{eq:kerrsen2}) by setting $a=0$ and non-vanishing values of the electric charge $Q$. The ISCO parameters for $Q=0.25$, $Q=0.5$ and $Q=0.75$ are shown in tables \ref{table:Q=0.25}, \ref{table:Q=0.5} and \ref{table:Q=0.75}. For a constant value of the electric charge, the ISCO parameters decrease with a similar behavior as that in the Schwarzschild background, as shown in Figure \ref{fig:ISCOparametersQ}. 

\begin{figure}[htb!]
	\centering
	\includegraphics[width=1 \linewidth]{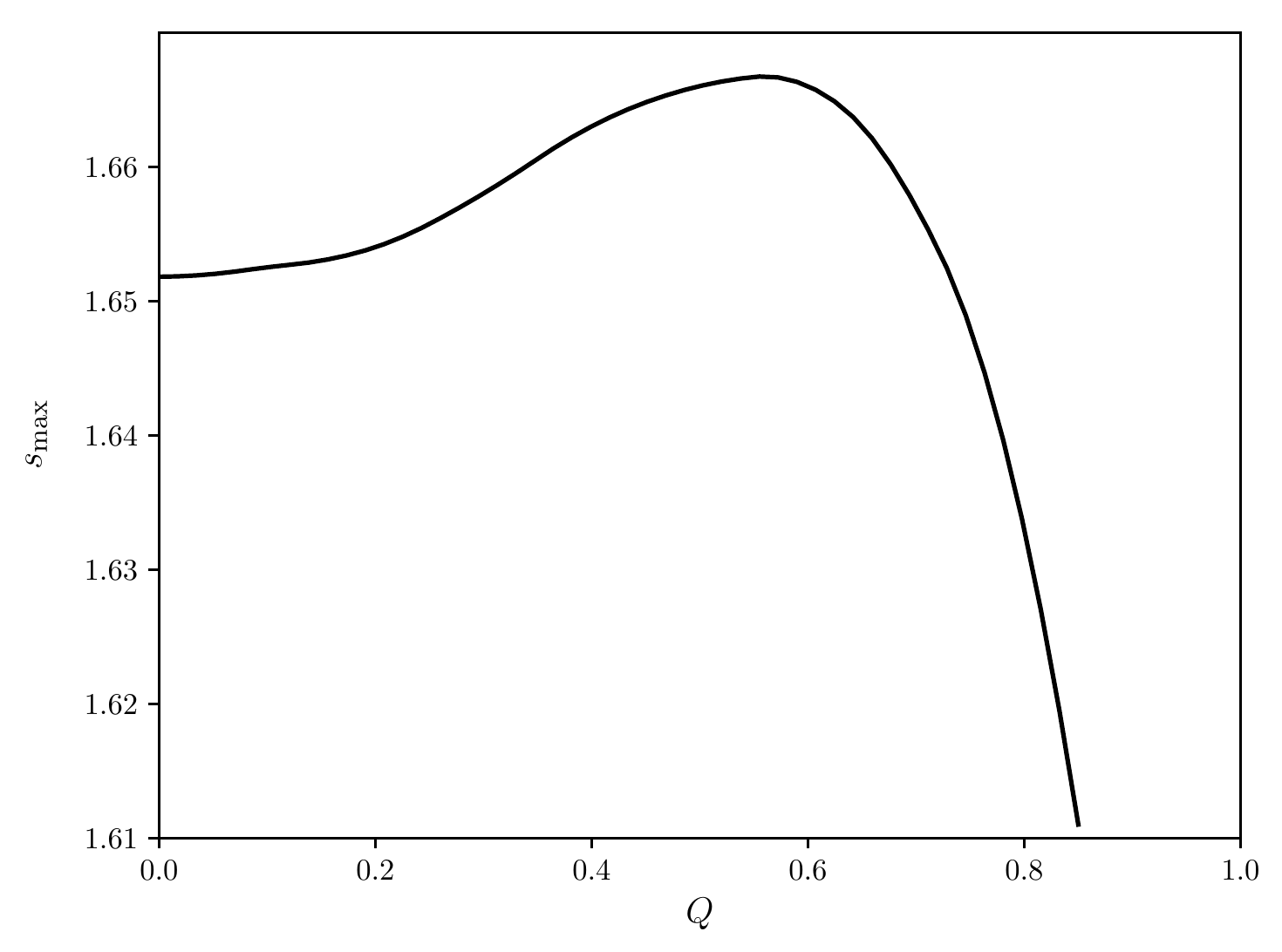}
	\caption{Behavior of the maximum spin $s_{max}$, given by the superluminal condition, as function of the electric charge, $Q$. } 
	\label{fig:smaxQ}
\end{figure}

\begin{figure}[htb!]
	\centering
	\includegraphics[width=1 \linewidth]{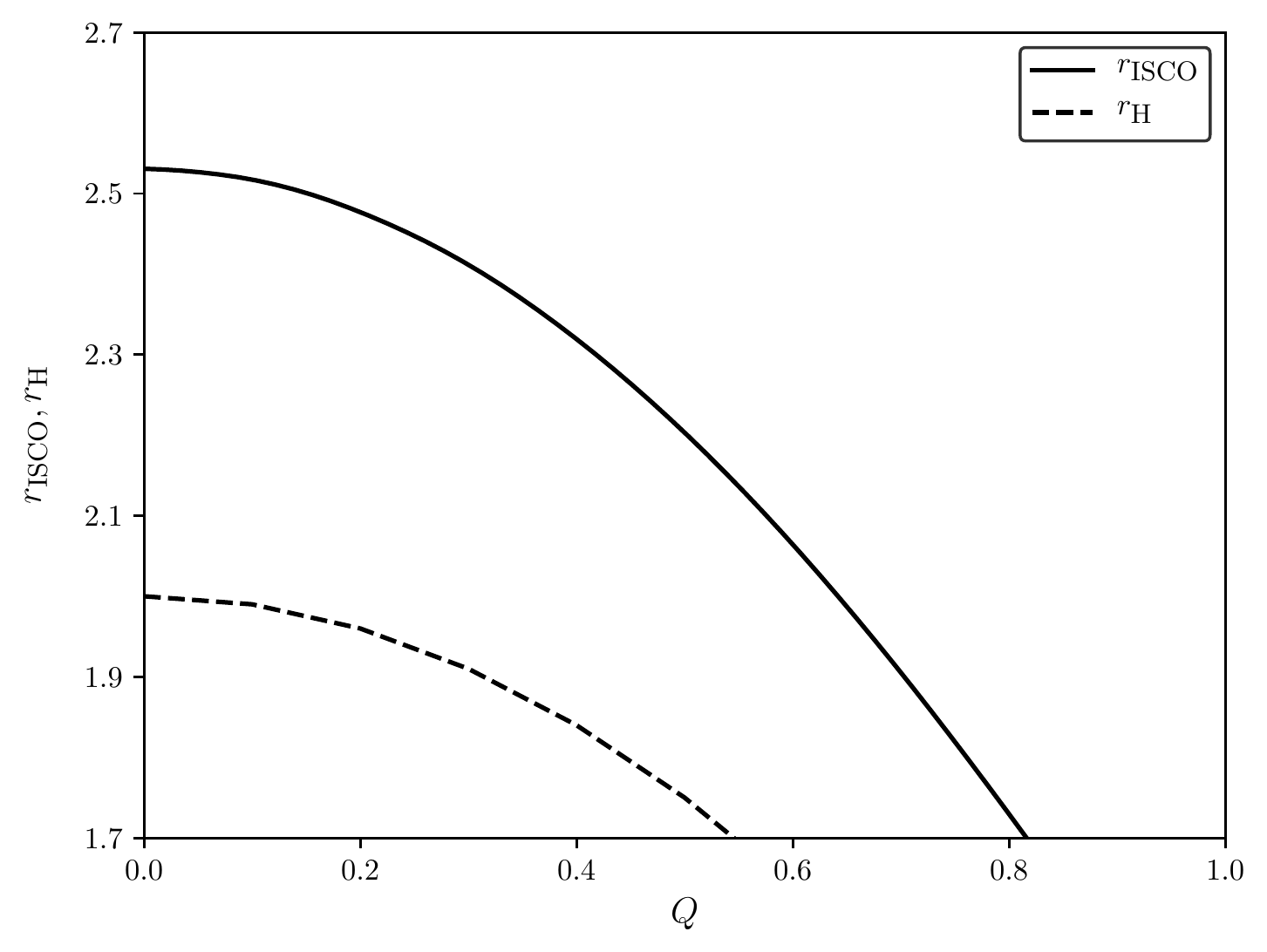}
	\caption{Radius of the ISCO for particles with the maximum spin $s_{max}$ moving in the GMGHS background, as function of the electric charge, $Q$. For comparison, it is also shown the radius of the event horizon,$r_H$, for the corresponding value of $Q$.} 
	\label{fig:rISCOmaxvsQ}
\end{figure}

This time, the superluminal constraint give maximum values of the particle’s spin depending on the given value of $Q$, with the behavior shown in Figure \ref{fig:smaxQ}. Note that there is a maximum in the curve around $Q = 0.55$. In Figure \ref{fig:rISCOmaxvsQ} it is shown the ISCO radius for particles moving with the maximum spin for each value of the electric charge and it is compared with the corresponding value of the event horizon's radius.  

\begin{table}[htp!]
\begin{tabular}{cccccc}
\hline
$s$    & $r_{ISCO}$    & $l_{ISCO}$    & $e_{ISCO}$    & $\Omega^{\frac{2}{3}}$ & $u^2 $ \\ \hline \hline
-0.9 	& 	7.0835 	& 	3.8271 	& 	0.9522	& 	0.1478 	& -0.5567 \\
-0.7 	&	6.8519 	& 	3.7516 	& 	0.9505	& 	0.1512 	& -0.5469 \\
-0.5 	&	6.6016 	& 	3.6703 	& 	0.9486  & 	0.1552 	& -0.5354 \\
-0.3 	& 	6.3300 	& 	3.5822 	& 	0.9463  & 	0.1599 	& -0.5217 \\
-0.1 	& 	6.0335 	& 	3.4858 	& 	0.9435  & 	0.1655 	& -0.5052 \\
0.0		& 	5.8746 	& 	3.4338 	& 	0.9420 	& 	0.1687 	& -0.4956 \\
0.1	    & 	5.7075 	& 	3.3789 	& 	0.9402 	& 	0.1723 	& -0.4849 \\ 
0.3 	& 	5.3458 	& 	3.2583	& 	0.9360  & 	0.1807 	& -0.4595 \\
0.5 	& 	4.9402 	& 	3.1192 	& 	0.9307 	& 	0.1915 	& -0.4267 \\ 
0.7 	& 	4.4815 	& 	2.9536 	& 	0.9233  & 	0.2056  & -0.3828 \\
0.9 	&	3.9660 	& 	2.7470 	& 	0.9124 	& 	0.2247 	& -0.3223 \\
\hline 
\end{tabular}
\caption{Parameters of the ISCO for a spinning particle moving in the GMGHS geometry with $a=0$ and $Q=0.25$.}
\label{table:Q=0.25}
\end{table}

\begin{table}[htp!]
\begin{tabular}{cccccc}
\hline
$s$    & $r_{ISCO}$    & $l_{ISCO}$    & $e_{ISCO}$    & $\Omega^{\frac{2}{3}}$ & $u^2 $ \\ \hline \hline
-0.9 	& 	6.6863 	& 	3.7433 	& 0.9504 	& 0.1551 	& -0.5385 \\ 
-0.7 	& 	6.4585 	& 	3.6662 	& 0.9485 	& 0.1584 	& -0.5298 \\ 
-0.5 	& 	6.2117 	& 	3.5830 	& 0.9464 	& 0.1622 	& -0.5193 \\ 
-0.3 	& 	5.9434 	& 	3.4926 	& 0.9439 	& 0.1667 	& -0.5065 \\
-0.1 	& 	5.6500 	& 	3.3934 	& 0.9410 	& 0.1722 	& -0.4907 \\ 
0. 		& 	5.4926 	& 	3.3398  & 0.9393 	& 0.1754 	& -0.4815 \\ 
0.1	    & 	5.3270 	& 	3.2830 	& 0.9374 	& 0.1790 	& -0.4711 \\ 
0.3 	& 	4.9684 	& 	3.1581 	& 0.9328 	& 0.1874 	& -0.4460 \\ 
0.5 	& 	4.5665 	& 	3.0132 	& 0.9268 	& 0.1984 	& -0.4131 \\ 
0.7 	& 	4.1140 	& 	2.8395 	& 0.9185 	& 0.2128 	& -0.3687 \\ 
0.9 	& 	3.6122 	& 	2.6217	& 0.9062 	& 0.2321 	& -0.3076 \\
\hline 
\end{tabular}
\caption{Parameters of the ISCO for a spinning particle moving in the GMGHS geometry with $a=0$ and $Q=0.5$}
\label{table:Q=0.5}
\end{table}

\begin{table}[htp!]
\begin{tabular}{cccccc}
\hline
$s$    & $r_{ISCO}$    & $l_{ISCO}$    & $e_{ISCO}$    & $\Omega^{\frac{2}{3}}$ & $u^2 $ \\ \hline \hline
-0.9 	& 5.9960 	& 3.5937 	& 0.9466 	& 0.1694 	& -0.5038 \\ 
-0.7 	& 5.7759 	& 3.5136 	& 0.9445 	& 0.1723 	& -0.4973 \\ 
-0.5 	& 5.5364 	& 3.4269 	& 0.9421 	& 0.1758 	& -0.4887 \\ 
-0.3 	& 5.2749 	& 3.3323 	& 0.9393 	& 0.1801 	& -0.4775 \\ 
-0.1 	& 4.9881 	& 3.2278 	& 0.9358 	& 0.1855 	& -0.4630 \\ 
0. 		& 4.8338 	& 3.1711 	& 0.9338 	& 0.1886 	& -0.4543 \\ 
0.1 	& 4.6715 	& 3.1109 	& 0.9316 	& 0.1922 	& -0.4442 \\ 
0.3 	& 4.3196 	& 2.9775 	& 0.9261 	& 0.2009 	& -0.4194 \\ 
0.5 	& 3.9265 	& 2.8213 	& 0.9189 	& 0.2122 	& -0.3860 \\ 
0.7 	& 3.4884 	& 2.6320 	& 0.9086 	& 0.2273 	& -0.3403 \\ 
0.9 	& 3.0171 	& 2.3924 	& 0.8931 	& 0.2474 	& -0.2778 \\
\hline 
\end{tabular}
\caption{Parameters of the ISCO for a spinning particle moving in the GMGHS geometry with $a=0$ and $Q=0.75$}
\label{table:Q=0.75}
\end{table}

\vspace{0.5cm}

\begin{figure}[htb!]
	\centering
	\includegraphics[width=1 \linewidth]{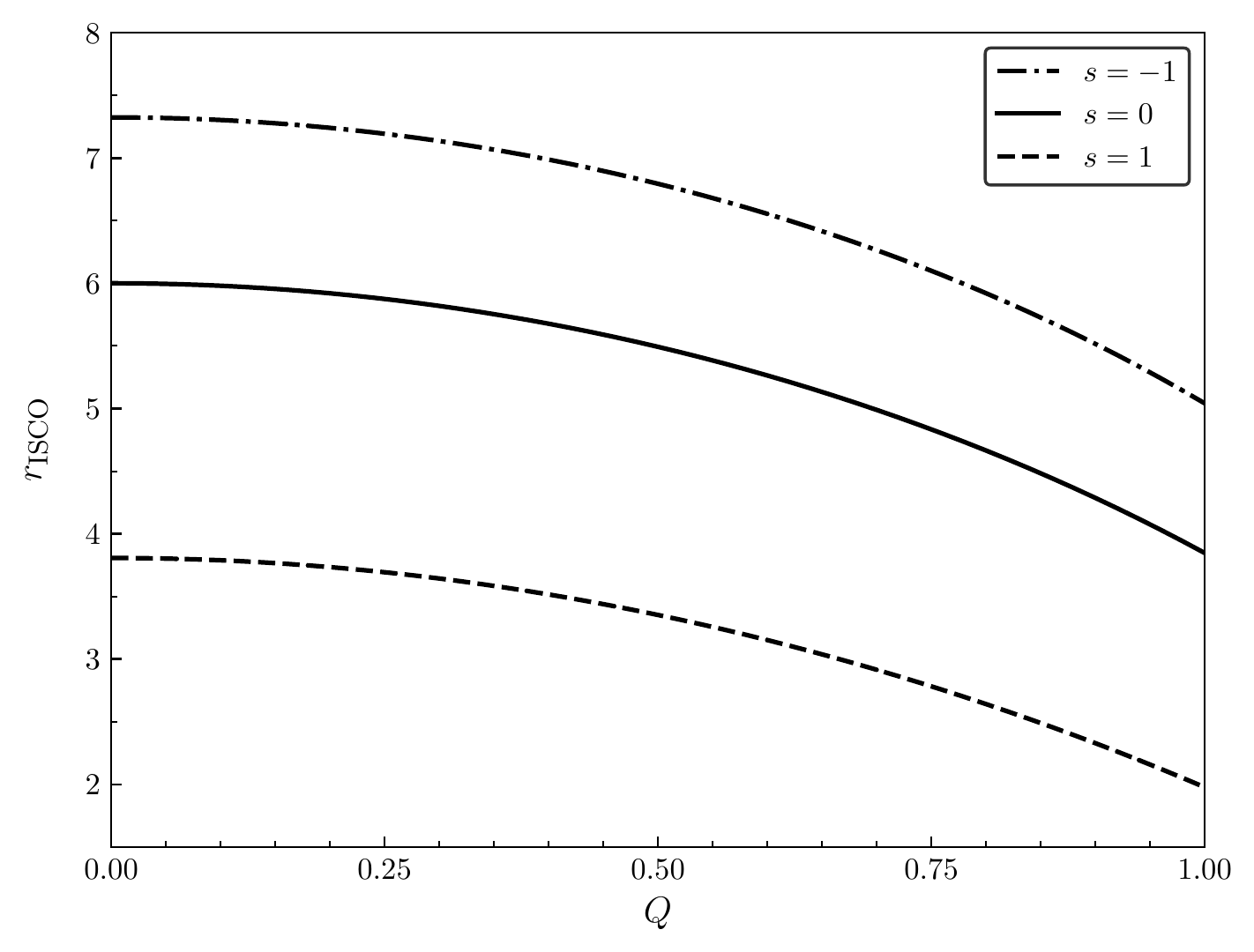}
	\caption{Relation between the radius of the ISCO of the spinning  test particle in the GMGHS background and the charge Q for three different values of the spin $s$.} 
	\label{fig:riscoQ}
\end{figure}

 On the other hand,  given a fixed value of the particle's spin, the radius of the ISCO decreases for increasing values of the electric charge. This is shown in Figure \ref{fig:riscoQ} for $s=-1$, $s=0$ and $s=1$. Notice that when $s=0$ the intercept is at $r_{\small \textrm{ISCO}}=6$ and the Schwarzschild case is recovered. Therefore, the overall effect of having a higher value for the black hole's electric charge is to decrease the radius of the ISCO. 

\subsection{Spinning Particle moving in a Kerr-Sen Background}
\begin{figure}[htb!]
	\centering
	\includegraphics[width=1 \linewidth]{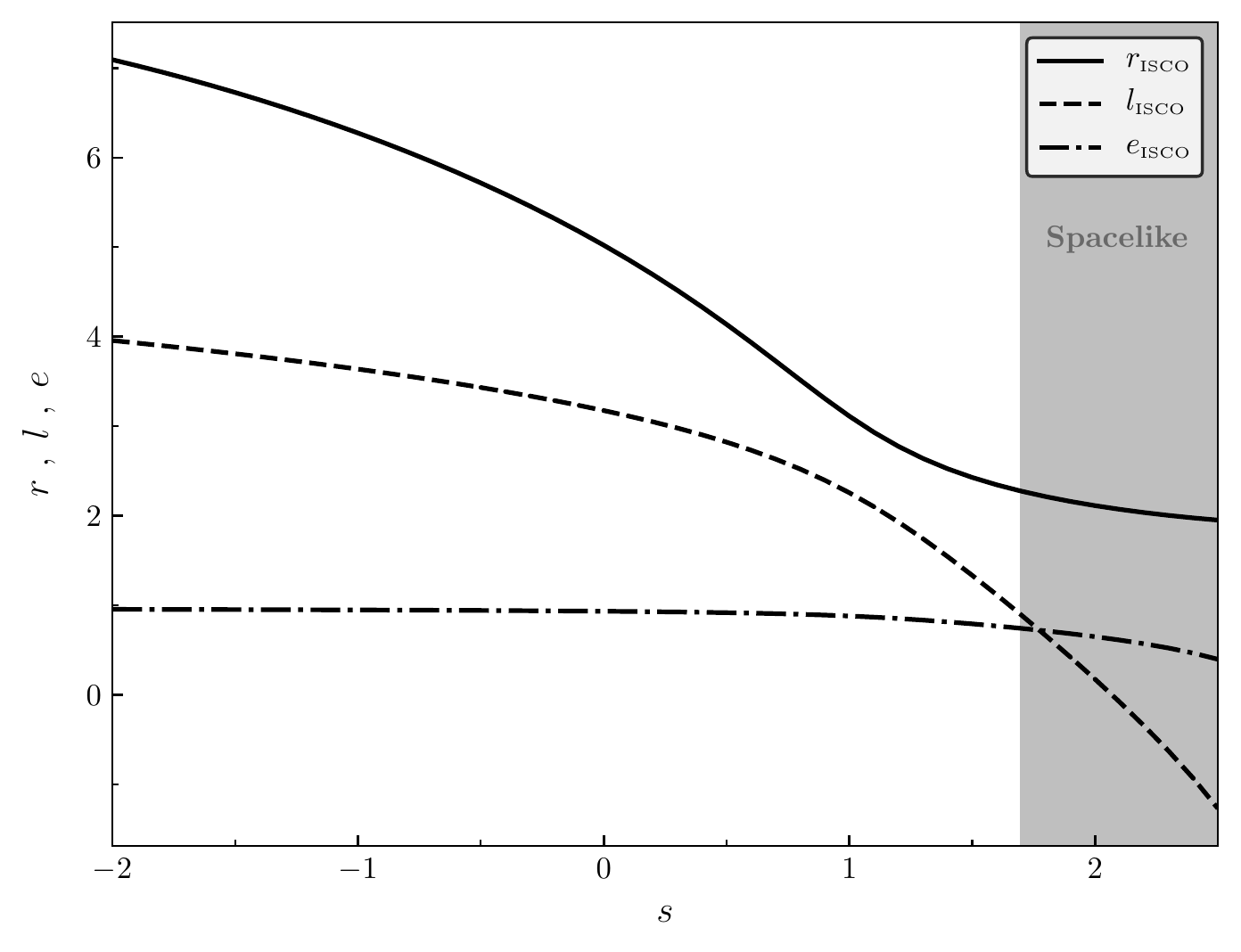}
	\caption{ISCO parameters in terms of the spin $s$ for a spinning test particle moving in the Kerr-Sen background with $a=0.25$ and $Q=0.25$. The superluminal constraint is located at $s_{\textrm{max}}=1.6959$. } 
	\label{fig:ISCOparametersaQ}
\end{figure}

Now we consider non-zero values for both parameters $a$ and $Q$. It is expected that the spin of the black hole must affect the motion the test particle and thus, modify the ISCO. The behavior is similar to that in Schwarzschild's and GMGHS backgrounds: when increasing the value of the particle's spin, the ISCO radius deecrases. However, there is another interesting result, the increase in the black hole's spin, $a$, makes the ISCO radius to decrease. The typical behavior of the ISCO parameters with the Kerr-Sen background is shown in Figure \ref{fig:ISCOparametersaQ} and some of the numerical results obtained for different combinations of the parameters $a$ and $Q$ are reported in Tables \ref{tab:ISCO-Sen-a0.25-Q0.25},  \ref{tab:ISCO-Sen-a0.5-Q0.5}, \ref{tab:ISCO-Sen-a0.5-Q0.75}, \ref{tab:ISCO-Sen-a0.75-Q0.25}.

When comparing our results with those obtained by Zhang et. al. \cite{Zhang2018} for the motion of the spinning particle in the Kerr-Newman background, we note a very important result: the ISCO radius in the Kerr-Sen background is always greater than the ISCO radius in the Kerr-Newman background with the same black hole's mass $M$, spin $a$ and electric charge $Q$ (for example compare our results with Tables III, IV and V in Ref. \cite{Zhang2018}). This fact may provide a method to discriminate one solution from the other from the observational point of view.

\begin{table}[]
\begin{tabular}{cccccc}
\hline
$s$    & $r_{ISCO}$    & $l_{ISCO}$    & $e_{ISCO}$    & $\Omega^{\frac{2}{3}}$ & $u^2 $ \\ \hline \hline
-0.9 	& 6.1729 	& 3.5991 	& 0.9454 	& 0.1685 	& -0.5059 \\
-0.7 	& 5.9548 	& 3.5188 	& 0.9432 	& 0.1726 	& -0.4950 \\
-0.5 	& 5.7174 	& 3.4318 	& 0.9406 	& 0.1774 	& -0.4820 \\
-0.3 	& 5.4583 	& 3.3367 	& 0.9376 	& 0.1832 	& -0.4662 \\
-0.1 	& 5.1743 	& 3.2316 	& 0.9340 	& 0.1902 	& -0.4469 \\
0. 		& 5.0220 	& 3.1745 	& 0.9318 	& 0.1943 	& -0.4357 \\
0.1 	& 4.8620 	& 3.1138 	& 0.9295 	& 0.1988 	& -0.4231 \\
0.3 	& 4.5173 	& 2.9791 	& 0.9237 	& 0.2095 	& -0.3931\\ 
0.5 	& 4.1376 	& 2.8216 	& 0.9160 	& 0.2231 	& -0.3549 \\
0.7 	& 3.7270 	& 2.6316 	& 0.9051 	& 0.2404	& -0.3057 \\
0.9 	& 3.3084 	& 2.3957 	& 0.8893 	& 0.2616 	& -0.2449 \\
\hline 
\end{tabular}
\caption{Parameters of the ISCO for a spinning particle moving in the Kerr-Sen geometry with $a=0.25$ and $Q=0.25$}
\label{tab:ISCO-Sen-a0.25-Q0.25}
\end{table}

\begin{table}[]
\begin{tabular}{cccccc}
\hline
$s$    & $r_{ISCO}$    & $l_{ISCO}$    & $e_{ISCO}$    & $\Omega^{\frac{2}{3}}$ & $u^2 $ \\ \hline \hline
-0.9 	& 4.6663 	& 3.2102 	& 0.9302 	& 0.2159 	& -0.3997 \\ 
-0.7 	& 4.4743 	& 3.1201 	& 0.92695 	& 0.2207 	& -0.3895 \\ 
-0.5 	& 4.2610 	& 3.0208 	& 0.9228 	& 0.2268 	& -0.3760 \\
-0.3 	& 4.0250 	& 2.9102 	& 0.9179 	& 0.2345 	& -0.3587 \\ 
-0.1 	& 3.7653 	& 2.7855 	& 0.9116 	& 0.2441 	& -0.3366 \\
0. 		& 3.6266 	& 2.7165 	& 0.9079 	& 0.2498 	& -0.3234 \\ 
0.1 	& 3.4823 	& 2.6424 	& 0.9036 	& 0.2562 	& -0.3086 \\ 
0.3 	& 3.1802 	& 2.4752 	& 0.8929 	& 0.2711 	& -0.2740 \\ 
0.5 	& 2.8719 	& 2.2765 	& 0.8782 	& 0.2890 	& -0.2326 \\ 
0.7 	& 2.5813 	& 2.0386 	& 0.8579 	& 0.3091 	& -0.1866 \\
0.9 	& 2.3336 	& 1.7576 	& 0.8306 	& 0.3297 	& -0.1404 \\
\hline 
\end{tabular}
\caption{Parameters of the ISCO for a spinning particle moving in the Kerr-Sen geometry with $a=0.5$ and $Q=0.5$}
\label{tab:ISCO-Sen-a0.5-Q0.5}
\end{table}

\begin{table}[]
\begin{tabular}{cccccc}
\hline
$s$    & $r_{ISCO}$    & $l_{ISCO}$    & $e_{ISCO}$    & $\Omega^{\frac{2}{3}}$ & $u^2 $ \\ \hline \hline
-0.9 	& 3.7403 	& 2.9765 	& 0.9186 	& 0.2551 	& -0.3219 \\ 
-0.7 	& 3.5671 	& 2.8792 	& 0.9143 	& 0.2595 	& -0.3157 \\ 
-0.5 	& 3.3700 	& 2.7706 	& 0.9089 	& 0.2658 	& -0.3051 \\ 
-0.3 	& 3.1492 	& 2.6480 	& 0.9021 	& 0.2743 	& -0.2894 \\ 
-0.1 	& 2.9062 	& 2.5074 	& 0.8934 	& 0.2853 	& -0.2681 \\ 
0. 		& 2.7776 	& 2.4289 	& 0.8880 	& 0.2919 	& -0.2550 \\ 
0.1 	& 2.6457 	& 2.3439 	& 0.8818 	& 0.2992 	& -0.2403 \\ 
0.3 	& 2.3797 	& 2.1510 	& 0.8660 	& 0.3161 	& -0.2066 \\ 
0.5 	& 2.1281 	& 1.9226 	& 0.8447 	& 0.3352 	& -0.1688 \\ 
0.7 	& 1.9116 	& 1.6558 	& 0.8165 	& 0.3550 	& -0.1308 \\ 
0.9 	& 1.7384 	& 1.3530 	& 0.7808 	& 0.3744 	& -0.0955 \\
\hline 
\end{tabular}
\caption{Parameters of the ISCO for a spinning particle moving in the Kerr-Sen geometry with $a=0.5$ and $Q=0.75$}
\label{tab:ISCO-Sen-a0.5-Q0.75}
\end{table}

\begin{table}[]
\begin{tabular}{cccccc}
\hline
$s$    & $r_{ISCO}$    & $l_{ISCO}$    & $e_{ISCO}$    & $\Omega^{\frac{2}{3}}$ & $u^2 $ \\ \hline \hline
-0.9 	& 3.9231 	& 2.9779 	& 0.9159 	& 0.2561 	& -0.3189 \\ 
-0.7 	& 3.7541 	& 2.8802 	& 0.9113 	& 0.2622 	& -0.3078 \\ 
-0.5 	& 3.5605 	& 2.7707 	& 0.9055 	& 0.2703 	& -0.2923 \\ 
-0.3 	& 3.3430 	& 2.6465 	& 0.8982 	& 0.2809 	& -0.2718 \\ 
-0.1 	& 3.1040 	& 2.5037 	& 0.8886 	& 0.2942 	& -0.2459 \\ 
0. 		& 2.9783 	& 2.4237 	& 0.8827 	& 0.3020 	& -0.2309 \\ 
0.1 	& 2.8503 	& 2.3369 	& 0.8759 	& 0.3106 	& -0.2145 \\ 
0.3 	& 2.5962 	& 2.1400 	& 0.8587 	& 0.3297 	& -0.1786 \\ 
0.5 	& 2.3631 	& 1.9075 	& 0.8354 	& 0.3504 	& -0.1412 \\ 
0.7 	& 2.1684 	& 1.6380 	& 0.8050 	& 0.3711 	& -0.1057 \\ 
0.9 	& 2.0152 	& 1.3340 	& 0.7669 	& 0.3910 	& -0.0741 \\
\hline 
\end{tabular}
\caption{Parameters of the ISCO for a spinning particle moving in the Kerr-Sen geometry with $a=0.75$ and $Q=0.25$}
\label{tab:ISCO-Sen-a0.75-Q0.25}
\end{table}

\section{Conclusion}

The study of the motion of a spinning test particle in the Kerr-Sen background was done by solving, numerically, the motion and spin evolution equations, obtaining the parameters defining the innermost stable circular orbit. We gave the ISCO parameters for different background cases, showing their dependence with the spin of the test particle. In summary this study yields the following results. The Schwarzschild's limit of the Kerr-Sen metric reproduces exactly the previously reported results by Zhang et. al. \cite{Zhang2018} and Jefremov et. al. \cite{Jefremov2015}. \\

When considering the motion of the particle in the GMGHS background, it is shown that the ISCO radius is always smaller than the obtained for the Schwarzschild's metric with the same black hole's mass $M$. However, when comparing the ISCO parameters with those obtained in the Reissner-Nördstrom background with the same mass and electric charge we find that the ISCO radius in the GMGHS metric is greater than the corresponding in the Reissner-Nördstrom metric \cite{Zhang2018}. \\

Finally, when considering the general case with non-vanishing $a$ and $Q$, we found that increasing spin of the test particle produces a decrease in the ISCO radius, consistently with the Schwarzschild's and GMGHS cases. When comparing the obtained ISCOs with those of a particle in the Kerr-Newman background \cite{Zhang2018}; it is shown that, in general, the ISCO radius in the Kerr-Sen background is greater than the ISCO radius in the Kerr-Newman case. \\

\emph{Acknowledgements}. This work was supported by the Universidad
Nacional de Colombia. Hermes Grant Code 41673.

\begin{widetext}
\section*{Appendix}

This appendix contains the necessary non-vanishing Riemann components to write the equations of motion \eqref{eq:EOM1} and \eqref{eq:EOM2} explicitly.

\begin{align}
R_{3003} &= \dfrac{\mu \Big( a^{2} - 2\mu r + r^{2} \Big)\Big( \mu a^{2}\cos^{2}\theta  \cosh^{2}\alpha - \mu a^{2}\cos^{2}\theta  \ +3a^{2}r \cos^{2}\theta - \mu r^{2}\cosh^{2}\alpha + \mu r^{2} - r^{3} \Big) \sin^{2}\theta \cosh^{2}\alpha}{ \left(  a^{2}\cos^{2}\theta + 2\mu r \cosh^{2}\alpha - 2\mu r + r^{2} \right)^{3} }
\end{align}

\begin{align}
\nonumber
R_{3113} &= \Big[\left( a^{2}-2\mu r +r^{2} \right) \left( 2\mu r \cosh^{2}\alpha + a^{2}\cos^{2}\theta -2\mu r + r^{2} \right)^{3}  \Big]^{-1} \bigg[ 3\mu \sin^{2}\theta \bigg[ -\dfrac{4}{3}\mu^{3}r^{2}\cosh^{8}\alpha  \left(a^{2}\cos^{2}\theta + a^{2} - 2\mu r \right) \\ \nonumber
& \ \  - \dfrac{8}{3}\mu^{2}r \cosh^{6}\alpha  \bigg( a^{2} \cos^{2}\theta \left( a^{2} - \dfrac{5}{2} \mu r  + \dfrac{5}{2}r^{2} \right) - 2a^{2}\mu r + 4\mu^{2}r^{2}  - \dfrac{3}{2}\mu r^{3} - \dfrac{1}{2}r^{4} \bigg) + \mu \left( a^{2}-2\mu r + r^{2} \right) \bigg( a^{4}\cos^{4}\theta \\ \nonumber
& \ \ + \cos^{2}\theta \Big( 6a^{2}\mu r -\dfrac{5}{3}a^{4} - 5a^{2}r^{2} \Big) + \dfrac{r^{2}}{3} \Big( a^{2}-24\mu^{2} + 6\mu r + 4r^{2} \Big) \bigg) \cosh^{4}\alpha  - \dfrac{2}{3}\cosh^{2}\alpha \left( a^{2} - 2\mu r + r^{2} \right) \\ \nonumber & \ \ \times  \bigg( a^{4}\cos^{4}\theta (\mu - 3r) - \dfrac{5}{2}a^{2}\cos^{2}\theta \left[ -r^{3} + \dfrac{11}{5}\mu r^{2} - r \left(\dfrac{9}{5}a^{2} + \dfrac{14}{5}\mu^{2} \right) + \mu a^{2}  \right] + \dfrac{1}{2}r^{2} \Big( \mu r^{2} + \left( 10\mu^{2}-3a^{2} \right)r  -r^{3} \\  & \ \ +\mu a^{2} - 16\mu^{3} \Big) \bigg) - \dfrac{1}{3}\mu \left(a^{2} - 2\mu r + r^{2} \right)\left( a^{2}\cos^{2}\theta - 2\mu r + r^{2}\right)^{2} \bigg] \bigg]
\end{align}

\begin{align}
\nonumber
R_{3101} &= -\Big[  \left( a^{2}-2\mu r +r^{2} \right) \left( 2\mu r \cosh^{2}\alpha + a^{2}\cos^{2}\theta -2\mu r + r^{2} \right)^{3} \Big]^{-1} \bigg[ \ 2\mu^{2}\cosh^{4}\alpha \left( a^{2}r \cos^{2}\theta - r^{3}  \right) \\ \nonumber & \ \ + \mu \cosh^{2}\alpha \left[ a^{2}\cos^{2}\theta \left( 5a^{2} - 12\mu r + 11r^{2} \right) - r^{2}\left( a^{2} - 4\mu r + 3r^{2} \right) \right] - \left( a^{2} - 2\mu r + r^{2}  \right) \Big( a^{2}\cos^{2}\theta (5\mu -9r)  \\  & \ \ -r^{2}(\mu - 3r) \Big) \bigg]  \mu a  \sin^{2}\theta \cosh^{2}\alpha 
\end{align}

\begin{align}
\nonumber
R_{1001} &= \Big[  \left( a^{2}-2\mu r +r^{2} \right) \left( 2\mu r \cosh^{2}\alpha + a^{2}\cos^{2}\theta -2\mu r + r^{2} \right)^{3} \Big]^{-1} \bigg[ \ \mu \cosh^{2}\alpha \bigg( a^{4}\cos^{4}\theta \left( \mu \cosh^{2}\alpha - \mu + 3r \right) \\ \nonumber 
& \ \ - a^{2}\cos^{2}\theta \Big[ \mu  \cosh^{2}\alpha \left( 5a^{2} -8\mu r + 5r^{2}\right) + 7r^{3} - 17\mu r^{2} + r \left( 9a^{2} + 8 \mu^{2}\right) - \mu a^{2} \Big] \\  & \ \ + r^{2} \left( \mu a^{2} \cosh^{2}\alpha - \mu a^{2} + 3a^{2}r - 4\mu r^{2} + 2r^{3}\right) \bigg) \bigg]
\end{align}

\end{widetext}

\end{document}